\documentclass[journal]{IEEEtran}

\usepackage{cite,algorithm,algorithmic,psfrag,bm,url,subfigure}
\usepackage[dvips]{graphicx}
\usepackage[cmex10]{amsmath}

\ifCLASSINFOpdf
\else
\fi

\begin{document}
%
\title{Investigation of Parameter Adaptation in RF Power Amplifier Behavioral Models}
%
%
%

\author{Ali~Soltani~Tehrani,
        Jessica~Chani,
        Thomas~Eriksson,
        and~Christian~Fager
\thanks{A. Soltani Tehrani, J. Chani and T. Eriksson are with the Department
of Signals and Systems, Chalmers University of Technology, Gothenburg,
Sweden e-mail: ali.soltani@me.com, thomase@chalmers.se.}
\thanks{C. Fager is with the Department of Microsystems and Nanotechnology, Chalmers University of Technology, Gothenburg,
Sweden e-mail: christian.fager@chalmers.se.}
}

\maketitle

\begin{abstract}
This paper presents an investigation into parameter adaptation in behavioral model--based digital predistortion for radio frequency power amplifiers. A novel measurement setup framework that emulates real--time adaptation in transmitters is developed that allows evaluation of different parameters, configurations and adaptation algorithms. This setup relieves the need for full feedback loops for parameter adaptation while providing the flexibility needed in the design process of parameter adaptation.

Issues such as convergence speed, sensitivity to quantization noise in the feedback loop and predistortion performance are investigated for some different parameter update algorithms using the proposed measurement setup. The approach presented in this paper allows the possibility to analyze different aspects of digital predistortion adaptation algorithms, and is an important enabling step for further research on parameter adaptation before the real--time hardware is implemented for use.
\end{abstract}


\IEEEpeerreviewmaketitle

\section{Introduction}
Current and future generation mobile communication systems require highly linear and power efficient transmitters to cope with the increasing demands on network speed and low power consumption. The tradeoff of high power efficiency and linearity has been well studied by researchers in the past few years with the goal of developing linearization techniques for nonlinear power amplifiers (PAs) with low power assumption. Digital predistortion (DPD) has established itself as a suitable candidate to compensate for PA nonlinearities while adding relatively modest additional power overhead to the transmitter architecture. Different DPD structures have been proposed, such as look--up--tables (LUTs) \cite{cavers1990,hammi07}, neural networks \cite{benvenuto93,ibnkahla2000} and Volterra--series based behavioral model DPDs \cite{pedro,isaksson,soltani}. Among them, Volterra--series based DPDs have been shown to be better suited for distortion compensation in modern wideband communication signals \cite{liu}.

In practical scenarios, to compensate for varying conditions in the transmitter such as PA aging, bias network variations, temperature shifts, and etc., parameter adaptation has been used \cite{cavers1990}. As communication systems move towards packet--based systems, and communication networks utilize techniques such as switching PAs off to conserve energy, rapid changes in PA input signal power and temperature drifts require more advanced adaptation systems. This has resulted in the need for complete feedback chains to construct a closed loop for adaptation in the transmitter. Such feedback chains however, greatly increase the hardware complexity in the transmitter and it is important to develop efficient, fast converging and as low-hardware-demanding as possible algorithms.

To construct a closed loop adaptation setup considerable resources are required, both in time and cost. Once this setup is constructed further, we are constrained to the designed setup and investigating different aspects becomes impractical. In order to implement and analyze adaptation algorithms in practice, due to the high cost in equipment and time required for designing a complete closed loop system, two approaches have been taken. One approach has been to develop adaptation techniques with LUT--based DPDs in either the direct path \cite{faulkner,chung2007,gilabert2008,presti}, or both the direct and feedback paths \cite{boo2009,woo2007,kim2010}. These approaches suffer from the huge increase in size of LUTs needed to compensate for the different drifts in PA behavior. Another approach has been to use neural networks and Volterra--series based structures and utilize iterative block--based updates \cite{rawat,liu,braithwaite2008,braithwaite2012}. However the performance of the different adaptation algorithms proposed in these works are evaluated using a approach where data blocks are uploaded consecutively to the measurement setup. This process is not consistent with closed loop adaptation systems, as the state of the PA will have changed while new data is uploaded at each block.

In this work, in order to investigate the performance of adaptation algorithms and evaluate both adaptation algorithms and how the feedback look should be designed, a measurement testbed is developed that mimics closed--loop adaptation with an open--loop setup. The flexibility of this setup enables the possibility to investigate different configurations for evaluation of parameter adaptation performance in Volterra--based DPDs. The setup is used to evaluate the performance of an indirect learning architecture (ILA) and a proactive adaptation technique proposed in \cite{soltaniims}. Issues such as convergence speed and the effect of quantization noise on the performance is investigated for these algorithms. Finally the testbed is used to investigate a new adaptation technique that reduces the sensitivity to noise in the feedback loop.


This paper is organized as follows. In Section II the main challenges in parameter adaptation is presented. In Section III an open loop adaptation testbed that mimics real--time closed loop systems is proposed. Section IV presents the measurement setup, devices and input data used and in Section V different adaptation algorithms are investigated, and a new adaptation technique is proposed. Finally conclusions are drawn in Section VI.

\section{Challenges in parameter adaptation}
\begin{figure}
\centering
\psfrag{a}[c][c][0.9]{DPD} \psfrag{b}[c][c][0.8]{DAC} \psfrag{c}[c][c][0.75]{Modulator} \psfrag{d}[c][c][1]{PA} \psfrag{e}[c][c][0.8]{Direct path} \psfrag{f}[c][c][0.8]{Feedback chain} \psfrag{g}[c][c][0.75]{Demodulator} \psfrag{h}[c][c][0.8]{ADC} \psfrag{i}[c][c][0.75]{Adaptation} \psfrag{j}[c][c][0.75]{algorithm} \psfrag{k}[c][c][0.9]{Real--time} \psfrag{l}[c][c][0.9]{hardware}
\includegraphics[width=\columnwidth]{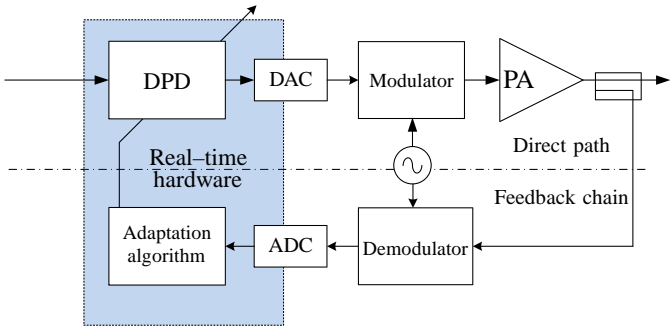}
\caption{A simple block diagram of a closed--loop transmitter chain for adaptive digital predistortion using Volterra--series based DPDs.}
\label{setup}
\end{figure}

In this section, a short background on important issues in adaptive parameter adaptation systems is presented in this section. A simple block diagram of a closed--loop adaptation system commonly employed in wireless transmitters is shown in Fig.~\ref{setup}. As shown in this figure, parameter adaptation of DPDs is commonly implemented with real--time hardware, which has high development costs both in terms of time and hardware. Further, commonly once these systems are designed, their settings are fixed and changing configurations and structures to analyze performance is highly time--consuming and expensive.

From Fig.~\ref{setup} it can be noticed that the performance of adaptive DPD systems is heavily dependent on both the adaptation algorithms used to update the parameters, and the quality of the signal in the feedback path. Issues such as bandwidth and quantization noise in the feedback loop, inphase and out of phase imbalance in the direct and feedback path, timing, convergence speed etc. are important issues that need to be analyzed and investigated, which requires a flexible measurement setup.

In Volterra--series based adaptive DPD literature, in order to investigate these and other issues the following  approach is commonly taken. A block of data is uploaded and captured from the setup (this block can either be the entire data set \cite{liu,braithwaite2012} or shorter blocks for faster updates \cite{braithwaite2008}), an adaptation technique, such as ILA or modified least squares (MLS), is used to update the parameters, then the next block of data (or the entire signal again) is passed through the adapted DPD and fed to the setup. Although the block based technique represents parameter adaptation, it is not able to accurately describe the closed loop adaptation that happen is practice where the PA is constantly run and not turned off for parameter updates. For example in traditional measurement setups, after the data is captured from the PA and fed to the PC, the PA receives no data while the new data is fed to the DPD and uploaded to the system, which means the the PA will face temperature drifts and the state of the PA is not consistent with the closed loop performance in real--time. In order to be able to mimic the real--time performance of adaptation systems, a measurement testbed needs to be constructed that addresses this issue.

\section{Adaptation testbed}
\begin{figure}
\centering
\psfrag{q}[c][c][0.75]{Initialization} \psfrag{w}[c][c][0.75]{Original data}
\psfrag{1}[c][c][0.55]{Block 1} \psfrag{2}[c][c][0.55]{Block 2} \psfrag{3}[c][c][0.55]{Block 3}
\psfrag{x}[c][c][0.6]{Initialization} \psfrag{v}[c][c][0.6]{step} \psfrag{y}[c][c][0.6]{Feedback} \psfrag{z}[c][c][0.6]{Adaptation} \psfrag{m}[c][c][0.6]{step 1} \psfrag{p}[c][c][1]{PA} \psfrag{b}[c][c][0.8]{Feedback} \psfrag{d}[c][c][0.8]{impairments} \psfrag{n}[c][c][0.6]{step 2}
\psfrag{c}[c][c][1]{DPD} \psfrag{u}[c][c][0.7]{update}
\includegraphics[width=0.98\columnwidth]{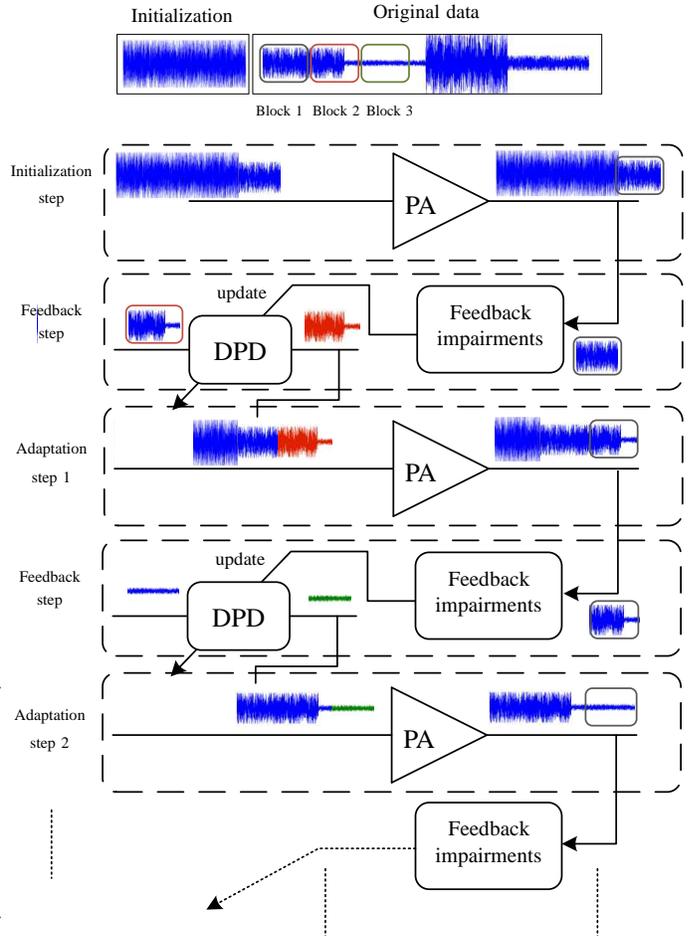}
\caption{Proposed open--loop testbed to mimic the closed loop adaptation DPD from Fig.~\ref{setup} without real--time hardware.}
\label{testbed}
\end{figure}

In order to accurately represent the closed loop adaptation transmitter in Fig.~\ref{setup} with an open--loop structure, we have to ensure that the PA is in a correct and consistent state with respect to the closed loop architecture. This is achieved by utilizing multiple measurements with overlapping blocks, where the final portion of the block is used at each step to update the DPD. By using enough of an overlap from the previous block we can ensure that the PA is in a correct state. The block diagram of the proposed open--loop adaptation testbed is shown in Fig.~\ref{testbed}.


In order to maintain a known state at the beginning of the process, an initialization block is added before the original data. The only requirement on this data is for it to be known, so the start of the data under analysis is not random. In the initialization step, after capturing the data from the output of the PA and sending it to the PC, we can add artificial hardware impairments in the digital domain, such as quantization noise, bandwidth limitations, etc, as we see fit to investigate robustness of the different adaptation algorithms. This is shown in the figure with the feedback impairments block. It should be noticed that only the final portion of the data is used for analysis, shown with a gray box, to ensure that the PA is in a correct state.

\begin{figure*}
\centering
\psfrag{1}[c][c][0.75]{Measurement 0}\psfrag{2}[c][c][0.75]{Measurement 1}\psfrag{3}[c][c][0.75]{Measurement 2}\psfrag{4}[c][c][0.75]{Measurement $n$}\psfrag{b}[c][c][0.8]{\textbf{Initialization data}}\psfrag{g}[c][c][0.8]{\textbf{Original data}}\psfrag{c}[c][c][0.8]{Original unpredistorted data}\psfrag{d}[c][c][0.75]{Data window uploaded to setup} \psfrag{e}[c][c][0.8]{\textbf{Analysis}}\psfrag{f}[c][c][0.85]{\textbf{to DPD}}\psfrag{h}[c][c][0.8]{Saved}
\psfrag{s}[c][c][0.75]{Step size ($S$)}
\includegraphics[width=0.8\linewidth]{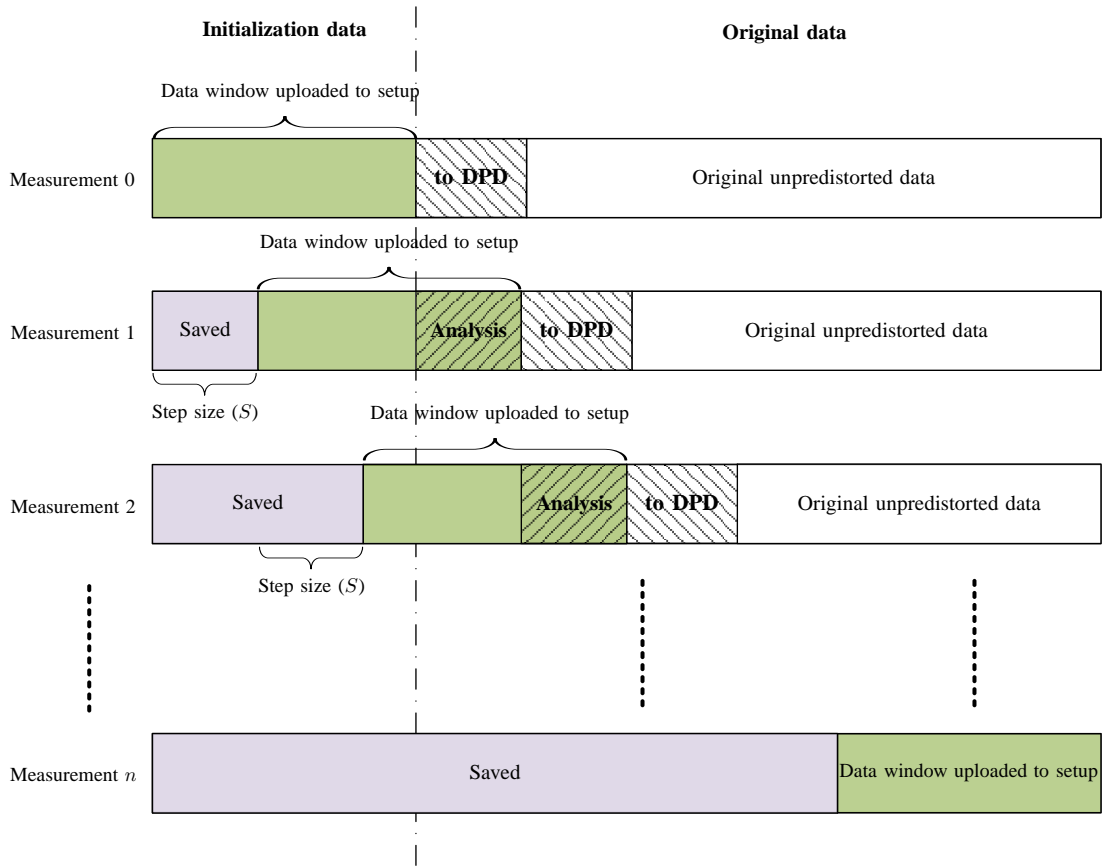}
\caption{The measurement setup used to emulate parameter adaptation in real--time. The white shaded \textbf{To DPD} block is the data portion that will be fed to the predistorter for the next round of measurements with the current set of parameters. The green box is the windowed block of data that is uploaded to the measurements system. This window includes a section to set the PA in the correct state at the beginning of the \textbf{To DPD} block, and the predistorted block. The analysis block (shown in the figure as \textbf{Analysis} and shaded) is used to update the parameters of the DPD after downloading the data from the setup for the next measurement. The data is uploaded with multiple measurements until the end of the data is reached.}
\label{repeatedmeasurements}
\end{figure*}

After adding artificial impairments to the signal, the captured PA output is used to update the parameters of the DPD using an adaptation technique. The next block of data is fed to the DPD with the updated parameters, and the output of the DPD is placed at the end of the block of data to be uploaded instead of the original block of data. This is shown for Adaptation step 1 with the red data used instead of the data labeled Block 2 and for step 3 with the green data.

The data to be uploaded to the proposed measurement setup at each measurement round is further shown in Fig.~\ref{repeatedmeasurements}. In this figure, it can be seen that through multiple measurements a windowed version of the data is uploaded to the setup, including a portion of the data that maintains a correct state at the beginning of the \textbf{to DPD} block. The window length and step sizes can be set by the user. By maintaining a large enough window size, the PA will be in a correct state for the new measurement, and the input/output data of that new measurement can be used to update the parameters of the DPD for the next measurement, in a manner that is consistent with a real--time adaptation setup.
%

\section{Measurement setup}
A short description of the measurement setup used in this work is provided in this section.
\subsection{Devices and instruments}
The modulator used is an Agilent E4438C vector signal generator (VSG) and an Agilent N9030A PXA signal analyzer is used as the vector signal analyzer to capture the data. The baseband I/Q data is generated in the computer and
downloaded to VSG. This signal is then fed to a 2.65 GHz $100$ W LDMOS Doherty power amplifier. The
PXA sends the RF signals back to the PC where they are
down-converted to baseband I/Q data. All devices are connected by
GPIB and all the data is time--aligned. In order to emulate real--time adaptation, the setup will be triggered manually a single time for each measurement, to ensure the PA is in the correct state.

\subsection{Data signals}
As the mobile usage pattern in current and future generation communication systems moves towards more bursty data signals \cite{soltaniims}, such dynamic traffic signals are the focus of this work. Use of such signals enables the evaluation of adaptation properties for different behavioral models in practical scenarios where PA state changes rapidly. The following two data signals are used.

\begin{itemize}
\item \emph{LTE test data}

A 20 MHz Long Term Evolution (LTE) test data E-UTRA Testmodel TM2 \cite{3gpp} is used for evaluation of modeling performances. A time record of the signal is shown in Fig.\ref{ltedata}. This signal is used to test the dynamic range of base stations, the error vector magnitude and the frequency error.
\begin{figure}
\centering
\includegraphics[width=0.85\columnwidth]{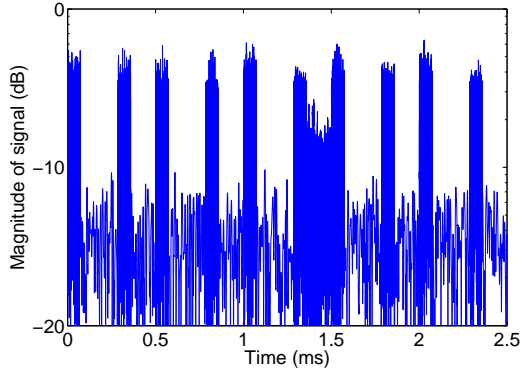}
\caption{Time record of the LTE data signal used.} \label{ltedata}
\end{figure}

\item \emph{Pulsed noise}

In order to mimic bursty usage patterns in future generation systems, while maintaining flexibility of choosing different configurations, the following settings from the 3rd Generation Partnership Project (3GPP) standard for LTE \cite{lte} are used to construct a pulsed input signal: 4 sub-frames (with 2ms for each) with a 10 dB power change. Fig.~\ref{input} shows the amplitude of the pulsed 4 MHz white Gaussian noise test signal.
\begin{figure}
\centering
\includegraphics[width=0.8\columnwidth]{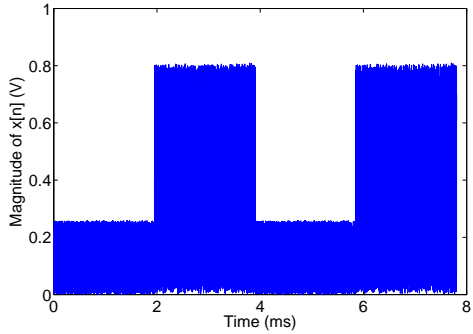}
\caption{The amplitude of the pulsed noise data signal.} \label{input}
\end{figure}
\end{itemize}


\section{Investigative Results}
In this section, we use the proposed testbed to investigate the performance for different adaptation algorithms with respect to issues such as convergence speed and sensitivity to quantization noise. A new algorithm is also proposed that has better noise handling properties. 

\subsection{Models and adaptation algorithms}
\begin{figure}
\centering
\psfrag{u}[c][c][0.85]{$u[n]$} \psfrag{x}[c][c][0.85]{$x[n]$} \psfrag{y}[c][c][0.85]{$y[n]$} \psfrag{e}[c][c][0.85]{$e[n]$} \psfrag{p}[c][c][0.85]{$x_{\text{post}}[n]$} \psfrag{a}[c][c][0.9]{Predistorter} \psfrag{d}[c][c][0.85]{(copy of $\mathbf{h}$)} \psfrag{c}[c][c][1]{PA} \psfrag{b}[c][c][0.9]{Postdistorter} \psfrag{f}[c][c][0.85]{(model $\mathbf{h}$)}
\includegraphics[width=0.95\columnwidth]{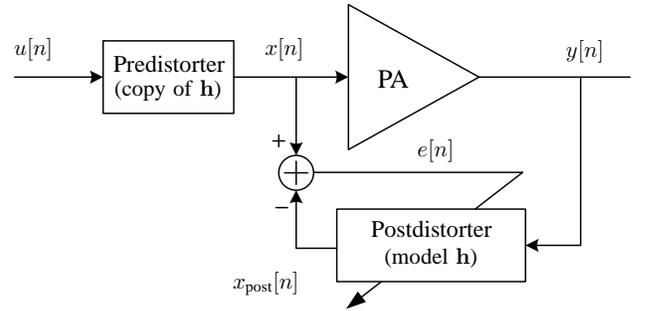}
\caption{Block diagram of the indirect learning architecture used for updating parameters}
\label{ila}
\end{figure}

Parameter adaptation techniques for Volterra--series based DPDs that are commonly employed in practice and literature follow the self-tuning controller structure \cite{paaso}. Both the direct and indirect learning architectures are for example, special cases of the self--tuning controller. In this work, the indirect learning architecture is employed for parameter adaptation. The block diagram of this technique is show in Fig.~\ref{ila}. 

In this structure, the output is fed to a post distorter and the error signal between the postdistorted value and the input to the PA is used to update the parameters, which are then copied to the predistorter. For Volterra--series models This can be written as
\begin{align}
\boldsymbol{\hat{\theta}} = (\mathbf{H}_y^H\mathbf{H}_y)^{-1}\mathbf{H}_y^H\mathbf{\text{x}}\\
\boldsymbol{\theta}_{\text{new}} = (1-\mu)\boldsymbol{\theta}_{\text{old}} + \mu\boldsymbol{\hat{\theta}}
\label{mulinear}
\end{align}
where $\mathbf{x}$ is the predistorted input to the PA, $\mu$ is the convergence constant (also called the forgetting factor), $\boldsymbol{\theta}_{\text{old}}$ is the parameter vector from the previous iteration,  $\boldsymbol{\theta}_{\text{new}}$ is the updated parameter vector, and $\mathbf{H}_y$ is the constructing matrix of the Volterra--series based model with the $\mathbf{y}$ (the block of data with the gray box in Fig.~\ref{testbed}). For example this matrix for a memory polynomial (MP) model \cite{kim} can be written as
\begin{equation}
\small
\mathbf{H}_{y[n]} =  \left[y[n]~~ y[n-1]~~ y[n]|y[n]|~~ y[n-1]|y[n-1]|~~ \cdots\ \right] \label{traditional}
\end{equation}
Any of the many Volterra--series based models developed in the literature can be used for adaptive DPD \cite{pedro}, and in this work the MP and generalized MP (GMP) \cite{morgan} models are used. It should be noticed that this structure should be used for models that are linear with respect to their parameters.

Another technique that is investigated in this work is with the proactive modeling technique proposed in \cite{soltaniims} where a state parameter is constructed that tracks long--term changes in the input signal. The model output can be written as
\begin{align}
y[n] = \mathbf{H}_{x[n]}\left(\bm{\theta}^{(0)} + s[n]\bm{\theta}^{(1)}\right), \label{model_structure}
\end{align}
where $s[n]$ is a state parameter of the PA, $\bm{\theta}^{(0)}$ are the parameters of the behavioral model independent of the state, $\bm{\theta}^{(1)}$ are called the dynamic parameters dependent on $s[n]$, and $\mathbf{H}_{x[n]}$ similar to (\ref{traditional}) are the same columns of common behavioral models (MP and GMP). It can be seen that in this technique while the parameters remain unchanged and are not updated, the output however is dependent on $s[n]\bm{\theta}^{(1)}$ which dynamically changes with a varying $s[n]$. In this work (similar to \cite{soltaniims}) the state parameter used is a low--passed version of $|x[n]|^2$.

In the next section different aspects of the proposed techniques are investigated with the help of the proposed testbed. For identification of the different techniques, unless otherwise specified, the DPD parameters are initialized from a separate measurement including the entire data set.

\subsection{Adaptive DPD performance}
\begin{figure}
\centering
\includegraphics[width=0.95\columnwidth]{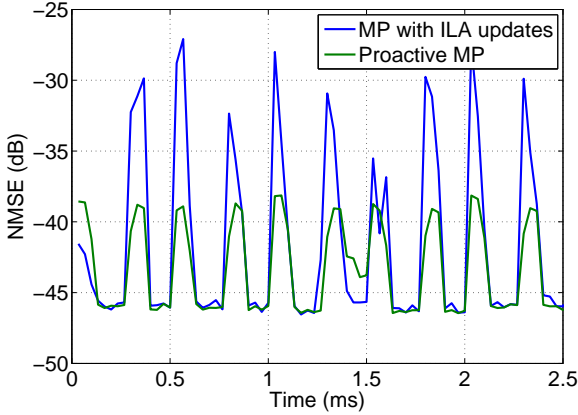}
\caption{NMSE vs time for the LTE data signal. The performance of the ILA adaptation of the reactive MP(7,4) with $\mu = 0.8$ is compared to the proactive MP(7,2).} \label{lte}
\end{figure}

Using the proposed measurement setup, and setting the window size to 120 thousand samples and the step size to 1024 samples (from Fig.~\ref{repeatedmeasurements}) the performance of the ILA parameter updates and the proactive model is investigated. To evaluate the performance at each iteration, the normalized mean square error (NMSE) is calculated for block of data at the end of the window (as shown with the gray box in Fig.~\ref{testbed}). The performance is shown for the LTE test data signal in Fig.~\ref{lte}. The model orders are chosen to maintain similar complexity between different techniques. It can be noticed that the reactive adaptation is too slow to react to the pulsed signal compared to the proactive model.

\begin{figure}
\centering
\includegraphics[width=0.95\columnwidth]{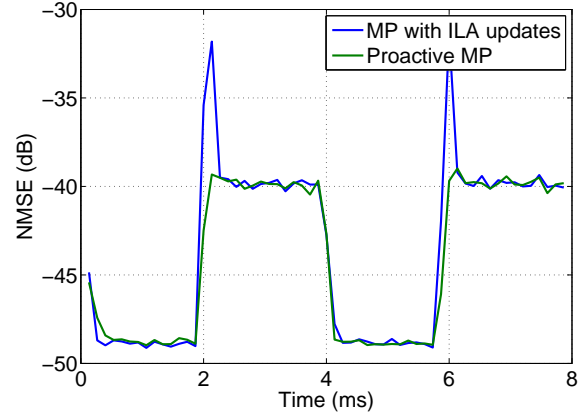}
\caption{NMSE vs time for the different models. In the case of the ILA $\mu=0.9$, MP(7,4) and for the proactive model MP(7,2).} \label{nmsetime}
\end{figure}

Fig.~\ref{nmsetime} shows the NMSE  vs time computed with a similar window size and step size of 4096 samples (equivalent to 0.13 ms) for the different models and the pulsed noise data signal. When the input data amplitude increases, it a loss in performance is noticed. In the case of MP model with ILA the parameter adaptation comes into effect and improves the NMSE. The same cycle can be noticed in the second iteration. For the proactive model however, it can be noticed that there is almost no loss in instantaneous performance during the switching to high power, as the model was able to proactively track the input signal power change.

\subsection{Convergence speed}
The speed of convergence in the parameters for adaptation algorithms is an important criterion for wireless transmitters. A fast converging algorithm results in less demand on the feedback loop, which can relax some of the hardware needed in the transmitter. In order to analyze the convergence speed two approaches can be taken. From (\ref{mulinear}) it can be noticed that $\mu$ can be varied to change the speed at which the parameters react to changes in the PA behavior. Another approach to analyze the convergence speed of parameter adaptation can be by varying the step size $(S)$ from Fig.~\ref{repeatedmeasurements}. A short step size represents fast updates required in the feedback loop, which can result in a better performance and quicker response to changes in the PA but also a dramatic increase in the hardware requirements for the feedback loop and, with smaller blocks, the parameter identification accuracy suffers.

\begin{figure}
\centering
\includegraphics[width=0.95\columnwidth]{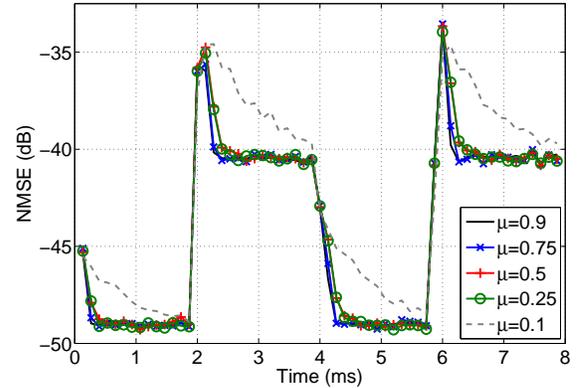}
\caption{Comparing the NMSE vs time for varying $\mu$ for the ILA update for an MP DPD.} \label{mufigure}
\end{figure}

In the first experiment the step size is kept constant and $\mu$ is varied to study the convergence performance of the ILA technique. Fig.~\ref{mufigure} shows the performance of the MP(7,4) model using ILA to update, with an analysis block size of 4096 samples. For this case the parameters were initialized with all parameters equal to zero except for the linear term. It can be seen that as $\mu$ increases the performance converges faster and for $\mu > 0.25$ the performance converges in around $0.3$ ms.

\begin{figure}
\centering
\includegraphics[width=0.95\columnwidth]{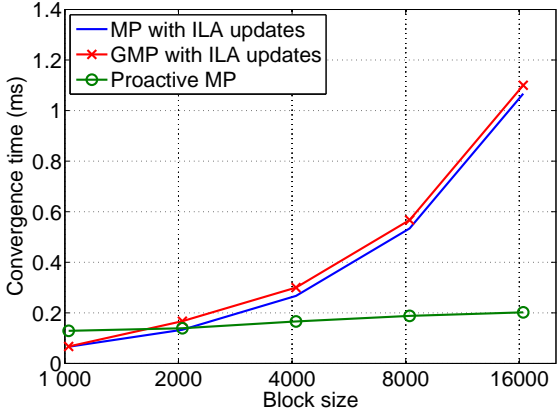}
\caption{Convergence vs analysis block length.} \label{blocklength}
\end{figure}

In the next experiment $\mu$ is kept constant at $0.8$ and the step size is varied. Fig.~\ref{blocklength} shows the convergence time for different models and algorithms vs the block length. It can be noticed that as the block length decreases (and the amount of updates from the feedback loop increase), the speed of convergence also decreases. In the case of the proactive model, it can be noticed that the parameter convergence remains constant over the block lengths. This type of adaptation, without requiring the extra hardware of the feedback loop to update the data, results in a convergence corresponding to the IDL method with a block length of around 2048 samples (corresponding to updating the parameters every 0.06 ms which may be demanding on the feedback loop hardware).

\subsection{Sensitivity to quantization noise}
In the experiment setup utilized in this work, we have used a high performance receiver. In practical scenarios in wireless transmitters, utilizing such a high performance receiver in the feedback path is too costly. In order to analyze the performance with a more practical receiver, in the feedback impairment block from Fig.~\ref{testbed}, a white Gaussian noise is added to the output signal (only for parameter updates, not for model evaluation) to represent quantization noise by having a less number of ADCs in the receiver. From \cite[p.~236]{datacompression}, it can be seen that the signal to quantization noise is proportional to
\begin{align}
\text{SNR} \propto 6.02n
\end{align}
where $n$ is the number of quantization bits.

\begin{figure}
\centering
\includegraphics[width=0.95\columnwidth]{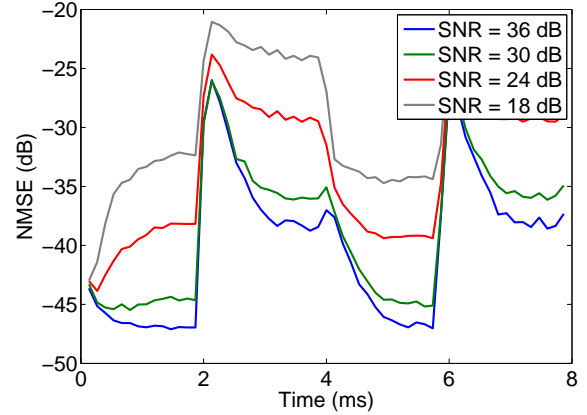}
\caption{Comparing the NMSE vs time for varying quantization noise (represented by signal to noise ratio in the feedback loop setting) for the ILA. For these results $\mu = 0.25$} \label{noise}
\end{figure}

Fig.~\ref{noise} shows the sensitivity of the an MP model with ILA updates to quantization noise in the feedback path.
It can be noticed that this updating algorithm is sensitive to the quality of the signal in the feedback loop, and for a signal to noise ratio (SNR) of less than 30 dB the performance degradation is severe.

\subsection{New algorithm with lower sensitivity to noise}
\begin{figure}
\centering
\psfrag{u}[c][c][0.9]{$u[n]$} \psfrag{x}[c][c][0.9]{$x[n]$} \psfrag{y}[c][c][0.9]{$y[n]$} \psfrag{e}[c][c][0.9]{$e[n]$} \psfrag{p}[c][c][0.9]{$x_{\text{post}}[n]$} \psfrag{a}[c][c][0.9]{Predistorter} \psfrag{m}[c][c][0.75]{(copy of $\mathbf{h}$)}\psfrag{b}[c][c][1]{PA} \psfrag{c}[c][c][0.9]{Postdistorter} \psfrag{d}[c][c][0.9]{Parameter} \psfrag{n}[c][c][0.9]{update} \psfrag{l}[c][c][0.75]{(model $\mathbf{h}$)} 
\includegraphics[width=0.95\columnwidth]{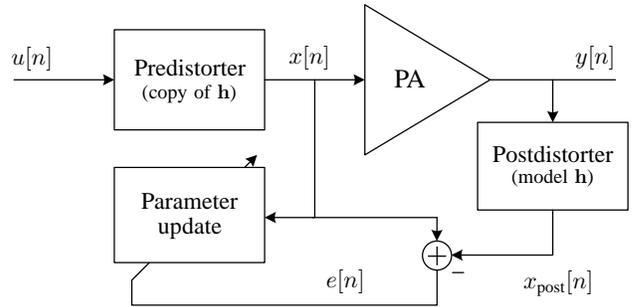}
\caption{Block diagram of proposed parameter update technique. The parameter updates are copied to both post and predistorter blocks.} \label{adaptnew}
\end{figure}

The high sensitivity to quantization noise for the ILA structure requires high performing ADC in the feedback loop, which are costly both in terms of hardware and in terms of power. In order to alleviate this issue an adaptation algorithm is proposed with the block diagram shown in Fig.~\ref{adaptnew}.

\begin{figure}
\centering
\includegraphics[width=0.95\columnwidth]{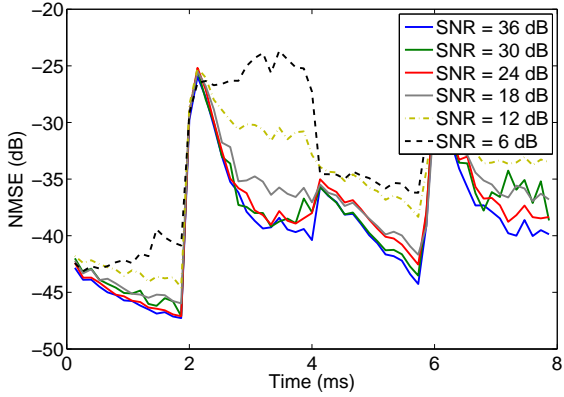}
\caption{NMSE vs time for varying quantization noise (represented by signal to noise ratio in the feedback loop setting) for the proposed update algorithm. For these results $\mu = 0.8$} \label{noise2m}
\end{figure}

The parameter updates for this technique are written as
\begin{equation}
\mathbf{x}_{\text{post}} = \mathbf{H}_y\theta_{\text{old}},
\end{equation}
\begin{equation}
\mathbf{e} = \mathbf{x}-\mathbf{x}_{\text{post}},
\end{equation}
\begin{equation}
\theta_{\text{new}} = \theta_{\text{old}} + \mu\left(\mathbf{H}_x^H\mathbf{H}_x\right)^{-1}\mathbf{H}_x^H\mathbf{e}. \label{munew}
\end{equation}
Using this technique, the PA input signal $\mathbf{x}$ (which is not affected by quantization noise) and the error signal are used to update the parameters. Fig.~\ref{noise2m} shows the convergence in performance for different quantization noise settings.

\begin{figure}
\centering
\includegraphics[width=0.85\columnwidth]{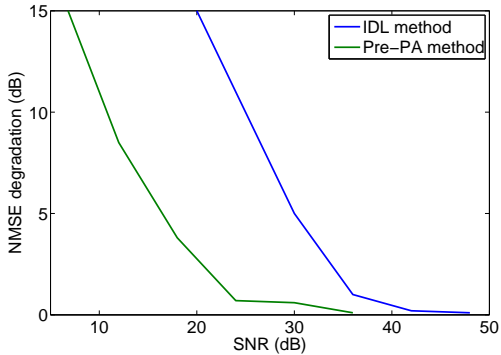}
\caption{Performance degradation vs the SNR in the feedback loop for the proposed parameter update technique.} \label{noise2}
\end{figure}

It can be noticed that compared to Fig.~\ref{noise}, the proposed technique is not as sensitive to noise in the feedback loop and for SNRs of up to 18 dB the performance is still acceptable. This is also observed from Fig.~\ref{noise2} where the performance degradation (defined as NMSE using the noisy output minus NMSE using noiseless output) for the proposed and ILA techniques are shown. The sharp performance degradation gives the minimum SNR needed to obtain good performance in parameter adaptation feedback loop hardware. It can be seen that the proposed technique has around 12-14 dB better performance compared to the ILA.

\begin{figure}
\centering
\includegraphics[width=0.85\columnwidth]{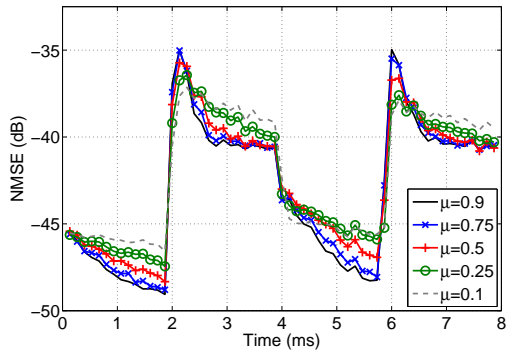}
\caption{NMSE vs time for varying $\mu$ with the proposed update technique.} \label{indir}
\end{figure}

The improved sensitivity to noise for this model however comes at the cost of convergence speed as seen from comparing  Fig.~\ref{indir} and Fig.~\ref{mufigure}. From Fig.~\ref{indir}, it can be noticed that the proposed technique is roughly 5 times slower to converge compared to ILA, which could be too slow for applications with rapid changes in input signal power.

\section{Conclusions}
In this paper an investigation into parameter adaptation for behavioral models used in digital predistortion is presented. A measurement setup testbed that mimics closed--loop real--time adaptation hardware with an open--loop system is developed. This setup enables a proper evaluation and investigation into parameter adaptation with Volterra--series based DPDs. The setup is used to investigate the performance of the indirect learning architecture in terms of the effect of quantization noise in the feedback loop, the convergence speed and the effect of the forgetting factor. A new update scheme is also proposed that is less sensitive to quantization noise in the feedback chain.

A proactive adaptation technique is also compared with the traditional technique on two data signals, a LTE test signal and a pulsed noise signal based on 3GPP standard. Results show that the proactive technique is able to achieve similar performance to the traditional technique without requiring the additional hardware in the feedback loop.

The framework proposed in this paper enables the analysis of parameter adaptation in behavioral--model based digital predistorters and provides an essential tool for future investigations of parameter adaptation.
\section*{Acknowledgement}
This research has been carried out in GigaHertz Centre in a joint project financed by the Swedish Governmental Agency
for Innovation Systems (VINNOVA), Chalmers University of Technology, Ericsson AB, Infineon Technologies Austria AG, NXP Semiconductors BV, and Saab AB.

\ifCLASSOPTIONcaptionsoff
  \newpage
\fi

\bibliographystyle{IEEEtran}
\bibliography{main}

\begin{thebibliography}{10}
\providecommand{\url}[1]{#1}
\csname url@samestyle\endcsname
\providecommand{\newblock}{\relax}
\providecommand{\bibinfo}[2]{#2}
\providecommand{\BIBentrySTDinterwordspacing}{\spaceskip=0pt\relax}
\providecommand{\BIBentryALTinterwordstretchfactor}{4}
\providecommand{\BIBentryALTinterwordspacing}{\spaceskip=\fontdimen2\font plus
\BIBentryALTinterwordstretchfactor\fontdimen3\font minus
  \fontdimen4\font\relax}
\providecommand{\BIBforeignlanguage}[2]{{%
\expandafter\ifx\csname l@#1\endcsname\relax
\typeout{** WARNING: IEEEtran.bst: No hyphenation pattern has been}%
\typeout{** loaded for the language `#1'. Using the pattern for}%
\typeout{** the default language instead.}%
\else
\language=\csname l@#1\endcsname
\fi
#2}}
\providecommand{\BIBdecl}{\relax}
\BIBdecl

\bibitem{cavers1990}
J.~Cavers, ``Amplifier linearization using a digital predistorter with fast
  adaptation and low memory requirements,'' \emph{Vehicular Technology, IEEE
  Transactions on}, vol.~39, no.~4, pp. 374 --382, nov 1990.

\bibitem{hammi07}
O.~Hammi, F.~Ghannouchi, S.~Boumaiza, and B.~Vassilakis, ``A data-based nested
  lut model for rf power amplifiers exhibiting memory effects,''
  \emph{Microwave and Wireless Components Letters, IEEE}, vol.~17, no.~10, pp.
  712 --714, oct. 2007.

\bibitem{benvenuto93}
N.~Benvenuto, F.~Piazza, and A.~Uncini, ``A neural network approach to data
  predistortion with memory in digital radio systems,'' in
  \emph{Communications, 1993. ICC 93. Geneva. Technical Program, Conference
  Record, IEEE International Conference on}, vol.~1, may 1993, pp. 232 --236
  vol.1.

\bibitem{ibnkahla2000}
M.~Ibnkahla, ``Neural network predistortion technique for digital satellite
  communications,'' in \emph{Acoustics, Speech, and Signal Processing, 2000.
  ICASSP '00. Proceedings. 2000 IEEE International Conference on}, vol.~6,
  2000, pp. 3506 --3509 vol.6.

\bibitem{pedro}
J.~C. Pedro and S.~A. Maas, ``A comparative overview of microwave and wireless
  power-amplifier behavioral modeling approaches,'' \emph{IEEE Trans. Microw.
  Theory Tech.}, vol.~53, no.~4, pp. 1150--1163, 2005.

\bibitem{isaksson}
M.~Isaksson, D.~Wisell, and D.~Ronnow, ``A comparative analysis of behavioral
  models for {RF} power amplifiers,'' \emph{IEEE Trans. Microw. Theory Tech.},
  vol.~54, no.~1, pp. 348--359, 2006.

\bibitem{soltani}
A.~S.Tehrani, H.~Cao, S.~Afsardoost, T.~Eriksson, M.~Isaksson, and C.~Fager,
  ``A comparative analysis of the complexity/accuracy tradeoff in power
  amplifier behavioral models,'' \emph{IEEE Trans. Mircow. Theory Tech.},
  vol.~58, pp. 1510--1520, 2010.

\bibitem{liu}
Y.~Liu, W.~Chen, J.~Zhou, B.~Zhou, F.~Ghannouchi, and Y.~Liu, \emph{Microwave
  Theory and Techniques, IEEE Transactions on}.

\bibitem{faulkner}
M.~Faulkner and M.~Johansson, ``Adaptive linearization using
  predistortion-experimental results,'' \emph{Vehicular Technology, IEEE
  Transactions on}, vol.~43, no.~2, pp. 323 --332, may 1994.

\bibitem{chung2007}
S.~Chung, J.~Holloway, and J.~Dawson, ``Open-loop digital predistortion using
  cartesian feedback for adaptive rf power amplifier linearization,'' in
  \emph{Microwave Symposium, 2007. IEEE/MTT-S International}, june 2007, pp.
  1449 --1452.

\bibitem{gilabert2008}
P.~Gilabert, A.~Cesari, G.~Montoro, E.~Bertran, and J.-M. Dilhac,
  ``Multi-lookup table fpga implementation of an adaptive digital predistorter
  for linearizing rf power amplifiers with memory effects,'' \emph{Microwave
  Theory and Techniques, IEEE Transactions on}, vol.~56, no.~2, pp. 372 --384,
  feb. 2008.

\bibitem{presti}
C.~Presti, D.~Kimball, and P.~Asbeck, ``Closed-loop digital predistortion
  system with fast real-time adaptation applied to a handset wcdma pa module,''
  \emph{Microwave Theory and Techniques, IEEE Transactions on}, vol.~60, no.~3,
  pp. 604 --618, march 2012.

\bibitem{boo2009}
H.~Boo, S.~Chung, and J.~Dawson, ``Adaptive predistortion using a
  $\delta\sigma$ modulator for automatic inversion of power amplifier
  nonlinearity,'' \emph{Vehicular Technology, IEEE Transactions on}, vol.~56,
  no.~12, pp. 901 -- 905, Dec 2009.

\bibitem{woo2007}
Y.~Y. Woo, J.~Kim, J.~Yi, S.~Hong, I.~Kim, J.~Moon, and B.~Kim, ``Adaptive
  digital feedback predistortion technique for linearizing power amplifiers,''
  \emph{Microwave Theory and Techniques, IEEE Transactions on}, vol.~55, no.~5,
  pp. 932 --940, may 2007.

\bibitem{kim2010}
J.~Kim, C.~Park, J.~Moon, and B.~Kim, ``Analysis of adaptive digital feedback
  linearization techniques,'' \emph{Circuits and Systems I: Regular Papers,
  IEEE Transactions on}, vol.~57, no.~2, pp. 345 --354, feb. 2010.

\bibitem{rawat}
M.~Rawat, K.~Rawat, and F.~Ghannouchi, ``Adaptive digital predistortion of
  wireless power amplifiers/transmitters using dynamic real-valued focused
  time-delay line neural networks,'' \emph{Microwave Theory and Techniques,
  IEEE Transactions on}, vol.~58, no.~1, pp. 95 --104, jan. 2010.

\bibitem{braithwaite2008}
R.~Braithwaite, ``Wide bandwidth adaptive digital predistortion of power
  amplifiers using reduced order memory correction,'' in \emph{Microwave
  Symposium Digest, 2008 IEEE MTT-S International}, june 2008, pp. 1517 --1520.

\bibitem{braithwaite2012}
R.~N. Braithwaite, ``Reducing estimator biases due to equalization errors in
  adaptive digital predistortion systems for rf power amplifiers,'' in
  \emph{Microwave Symposium Digest (MTT), 2012 IEEE MTT-S International}, june
  2012.

\bibitem{soltaniims}
A.~S. Tehrani, T.~Eriksson, and C.~Fager, ``Modeling of long term memory
  effects in rf power amplifiers with dynamic parameters,'' in \emph{Microwave
  Symposium Digest (MTT), 2012 IEEE MTT-S International}, june 2012, pp. 1 --3.

\bibitem{3gpp}
\BIBentryALTinterwordspacing
ETSI. (2006, Dec.) Lte; evolved universal terrestrial radio access (e-utra);
  base station (bs) conformance testing (3gpp ts 36.141 version 8.3.0 release
  8). [Online]. Available:
  \url{http://www.etsi.org/deliver/etsi_ts/136100_136199/136141/08.03.00_60/ts_136141v080300p.pdf}
\BIBentrySTDinterwordspacing

\bibitem{lte}
\BIBentryALTinterwordspacing
J.~Zyren. (2007, Aug,) Overview of the 3gpp long term evolution physical layer.
  [Online]. Available:
  \url{http://www.freescale.com/files/wirelesscomm/doc/whitepaper/3GPPEVOLUTIONWP.pdf}
\BIBentrySTDinterwordspacing

\bibitem{paaso}
H.~Paaso and A.~Mammela, ``Comparison of direct learning and indirect learning
  predistortion architectures,'' in \emph{Wireless Communication Systems. 2008.
  ISWCS '08. IEEE International Symposium on}, oct. 2008, pp. 309 --313.

\bibitem{kim}
J.~Kim and K.~Konstantinou, ``Digital predistortion of wideband signals based
  on power amplifier model with memory,'' \emph{Electron. Lett.}, vol.~27, pp.
  1417--1418, 2001.

\bibitem{morgan}
D.~Morgan, Z.~Ma, J.~Kim, M.~Zierdt, and J.~Pastalan, ``A generalized memory
  polynomial model for digital predistortion of rf power amplifiers,''
  \emph{IEEE Trans. Sig. Proc.}, vol.~54, pp. 3852 --3860, oct. 2006.

\bibitem{datacompression}
K.~Sayood, \emph{Introduction to data compression, 3rd ed.}\hskip 1em plus
  0.5em minus 0.4em\relax Morgan Kaufmann, 2006.

\end{thebibliography}

\end{document}